\documentclass[3p]{elsarticle}

\usepackage{bbm, bm, epsfig, amsmath, amsfonts, amssymb, wasysym, graphicx}
\usepackage{hyperref}

\journal{CNSNS}
\begin{document}

\begin{frontmatter}
\title{Boltzmann equation simulation for a trapped Fermi gas of atoms}

\author[damtp,lmu]{O. Goulko\corref{cor1}}
\address[damtp]{Department of Applied Mathematics and Theoretical Physics, University of Cambridge, Centre for Mathematical Sciences, Cambridge, CB3 0WA, United Kingdom}
\address[lmu]{Present address: Physics Department, Arnold Sommerfeld Center for Theoretical Physics, and Center for NanoScience, Ludwig-Maximilians-Universit\"at, Theresienstra\ss e 37, 80333 Munich, Germany}
\ead{O.Goulko@physik.uni-muenchen.de}

\author[ens]{F. Chevy}
\address[ens]{Laboratoire Kastler Brossel, Ecole Normale Sup\'erieure, CNRS, UPMC, 24 rue Lhomond, 75231 Paris Cedex 05, France}
\ead{frederic.chevy@lkb.ens.fr}

\author[sh]{C. Lobo}
\address[sh]{School of Mathematics, University of Southampton, Highfield, Southampton, SO17 1BJ, United Kingdom}
\ead{C.Lobo@soton.ac.uk}

\date{\today}

\begin{abstract}
The dynamics of an interacting Fermi gas of atoms at sufficiently high temperatures can be efficiently studied via a numerical simulation of the Boltzmann equation. In this work we describe in detail the setup we used recently to study the oscillations of two spin-polarised fermionic clouds in a trap. We focus here on the evaluation of interparticle interactions. We compare different ways of choosing the phase space coordinates of a pair of atoms after a successful collision and demonstrate that the exact microscopic setup has no influence on the macroscopic outcome.
\end{abstract}

\begin{keyword}
Boltzmann equation \sep cold atoms \sep interacting Fermi gas
\PACS 03.75.Ss \sep 03.75.Hh
\end{keyword}

\end{frontmatter}

\relpenalty=9999
\binoppenalty=9999

\newcommand{\Ei}{\textnormal{Ei}}
\newcommand{\Li}{\textnormal{Li}}
\newcommand{\fermiT}{\tilde{T}_{\rm F}}
\newcommand{\fermien}{\tilde{\varepsilon}_{\rm F}}
\newcommand{\fermimom}{\tilde{k}_{\rm F}}

\section{Introduction}
The Boltzmann simulation of the semiclassical regime of trapped Fermi gases \cite{inguscio2006ultracold} has become a crucial tool to understand recent experiments \cite{Zwierlein, ZwierleinImbalanced, Trenkwalder, Thomas} where spin-polarised atomic clouds undergo strong collisions. The simulation is one of the very few methods which allow us to calculate accurately e.g.\ spin transport coefficients in ways that can then be quantitatively compared with experiment. In a previous work \cite{boltzmannletter} we studied the collision of two spin-polarised fermionic clouds and obtained excellent agreement with experimental results. The goal of the present paper is to describe in detail the numerical setup that was used there and to present tests of the simulation and other information which is potentially very useful for those wishing to implement this important method.

Our setup is related to that of \cite{urban} but we introduce several new modifications \cite{mythesis}. For instance, we place an artificial grid over the continuous coordinate space that allows us to evaluate particle collisions efficiently. However, the main difference between this work and \cite{urban} is the method used to choose the new positions and momenta of two colliding particles after the collision. A quantum mechanical collision is a fundamentally random process with constraints given by the symmetries of the system. There are several ways to implement these constraints within the framework of a molecular dynamics simulation. We will discuss the physics of several collisional setups and also compare the numerical results associated with them in detail. We find that, although they exhibit very similar macroscopic behaviour, there are distinct technical advantages which make some of the methods preferable over others.

In Sec.~\ref{sec:numsetup} of this paper we introduce the general numerical setup. We describe the method of test particles and explain what conditions need to be fulfilled for a semiclassical collision to take place and how to efficiently search the system for suitable collision partners. The different ways of choosing the new positions and momenta after a collision are discussed in detail in Sec.~\ref{sec:collsetup}. In Sec.~\ref{sec:boltzmanntests} we show how to determine the optimal simulation parameters and present several tests of the numerical method, in particular also of the different collisional setups. A summary of our results and an outlook on future work can be found in Sec.~\ref{sec:summary}.

\section{Numerical setup}
\label{sec:numsetup}
\subsection{Definitions}
We consider a system of two-component fermions with equal mass $m$, labelled by the spin index $s=\{\uparrow,\downarrow\}$ and work in units in which $\hbar=k_{\rm B}=1$. Fermions of opposite spin can interact via $s$-wave scattering and the cross section is given by
\begin{equation}
\sigma=\frac{4\pi a^2}{1+a^2\mathbf{p}_\textnormal{rel}^2/4},
\end{equation}
where $a$ is the scattering length and $\mathbf{p}_\textnormal{rel}=\mathbf{p}_\uparrow-\mathbf{p}_\downarrow$ is the relative momentum of the two atoms. We assume that the system is in the normal phase and that the temperature is sufficiently high, so that the two spin distributions can be described semiclassically in terms of functions $f_s(\mathbf{r},\mathbf{p},t)$. Within the local density approximation the equilibrium distribution for fermions is given by the Fermi-Dirac distribution
\begin{equation}
f_s^{(\textnormal{FD})}(\mathbf{r},\mathbf{p})=\frac{1}{e^{(p^2/2m+V(\mathbf{r})-\mu_s)/T}+1},
\label{eqn:fermidirac}
\end{equation}
where $T$ is the temperature and $\mu_s$ the chemical potential defined by the normalisation condition
\begin{equation}
N_s=N/2=\int d^3r n_s(\mathbf{r},t)=\int d^3r\int \frac{d^3p}{(2\pi)^3} f_s(\mathbf{r},\mathbf{p},t).
\label{FDnorm}
\end{equation}
The atom number for each species is assumed to be equal, $N_\uparrow=N_\downarrow=N/2$, and will be held fixed during the simulation. The trapping potential is assumed to be harmonic,
\begin{equation}
V(\mathbf{r})=\frac{1}{2}m(\omega_x^2x^2+\omega_y^2y^2+\omega_z^2z^2),
\end{equation}
with the three trapping frequencies $\omega_x$, $\omega_y$ and $\omega_z$. In the following we will only consider the isotropic case $\omega_x=\omega_y=\omega_z$ or a cigar shaped trap with $\omega_z<\omega_x=\omega_y\equiv\omega_\perp$. We denote the geometric average of the trap frequencies as $\omega_0=(\omega_x\omega_y\omega_z)^{1/3}$.

We define the Fermi energy from the energy distribution of the non-interacting gas at zero temperature, $\fermien=\fermiT=(3N)^{1/3}\omega_0$. This energy scale marks the frontier between the classical and quantum regimes. For $T\gtrsim\fermiT$ the fermions behave essentially like classical particles and can be described by the Maxwell-Boltzmann distribution,
\begin{equation}
f_s^{(\textnormal{MB})}(\mathbf{r},\mathbf{p})=N_s\frac{\omega_0^3}{T^3} {\rm e}^{-(p^2/2m+V(\mathbf{r}))/T}.
\end{equation}
The Fermi energy can also be used to determine the typical scales of the cloud, $\fermien=\fermimom^2/2m=m\omega_i^2R_i^2/2$, where $\fermimom$ is the Fermi momentum and $R_i=\fermimom/(m\omega_i)$ are the Thomas-Fermi radii in the three spatial directions. These quantities give the widths of the zero-temperature momentum and density distributions respectively and are therefore useful to describe the extent of the cloud in momentum and coordinate space.

The time-evolution of the distribution function $f_s(\mathbf{r},\mathbf{p},t)$ is given by the Boltzmann equation,
\begin{equation}
\partial_t f_s+(\mathbf{p}/m)\cdot\boldsymbol{\nabla}_rf_s-\boldsymbol{\nabla}_rV\cdot\boldsymbol{\nabla}_pf_s=-I[f_s,f_{\overline{s}}],
\end{equation}
where the left-hand side represents the propagation of the atoms in the potential and the right-hand side stands for the collision integral, which depends on the particle statistics. Here, we consider only collisions between atoms carrying opposite spins. Indeed, for ultracold fermions, the Pauli principle forbids interactions between particles of the same spin \cite{inguscio2006ultracold}. At temperatures $T\gtrsim\fermiT$ when the Maxwell-Boltzmann distribution is applicable, the collision integral takes the form
\begin{equation}
I_{\rm class}[f_s,f_{\overline{s}}]=\int\frac{d^3p_{\overline{s}}}{(2\pi)^3}\int d\Omega\frac{d\sigma}{d\Omega}\frac{|\mathbf{p}_s-\mathbf{p}_{\overline{s}}|}{m}[f_sf_{\overline{s}}-f'_sf'_{\overline{s}}],
\label{classcollint}
\end{equation}
where the indices $s$ and $\overline{s}$ label the two colliding atoms of opposite spin, the primed variables refer to quantities after the collision and $\Omega$ is the angle between the incoming and outgoing relative momenta. The Pauli exclusion principle only allows fermions to scatter into a previously unoccupied quantum state. This reduces the scattering probability by the Pauli blocking term proportional to $(1-f'_s)(1-f'_{\overline{s}})$. Taking this into account, the collision integral reads
\begin{equation}
I[f_s,f_{\overline{s}}]=\int\frac{d^3p_{\overline{s}}}{(2\pi)^3}\int d\Omega\frac{d\sigma}{d\Omega}\frac{|\mathbf{p}_s-\mathbf{p}_{\overline{s}}|}{m}[f_sf_{\overline{s}}(1-f'_s)(1-f'_{\overline{s}})-f'_sf'_{\overline{s}}(1-f_s)(1-f_{\overline{s}})].
\label{qmcollint}
\end{equation}
At temperatures $T\gtrsim\fermiT$ the Pauli terms are close to one, and therefore the collision integral (\ref{qmcollint}) tends to the classical form (\ref{classcollint}).

\subsection{Test particles}
To efficiently simulate the evolution of the continuous distribution function $f_s(\mathbf{r},\mathbf{p},t)$ we use the method of test particles. These are point-like particles which form a discrete approximation to $f_s(\mathbf{r},\mathbf{p},t)$ through $\delta$-functions. In order for this approximation to be accurate we will represent each fermion by several test particles \cite{urban, jackson, wade}. The higher the ratio $\tilde{N}/N$ of test particles to atoms, the more precisely the continuous distribution function will be approximated,
\begin{equation}
f_s(\mathbf{r},\mathbf{p},t)\rightarrow\frac{N_s}{\tilde{N}_s}\sum_{i=1}^{\tilde{N}_s}(2\pi)^3\delta(\mathbf{r}-\mathbf{r}_i(t))\delta(\mathbf{p}-\mathbf{p}_i(t)).
\label{eq:fsdelta}
\end{equation}
Physical observables need to be rescaled, for instance the test particle cross section becomes $\tilde{\sigma}=\sigma(N/\tilde{N})$. A generic thermal expectation value of a single-particle observable $X(\mathbf{r},\mathbf{p})$ can be easily calculated within the test particle picture, as the integration reduces to a sum over all test particles,
\begin{eqnarray}
\langle X\rangle&=&\sum_s\frac{1}{N_s}\int d^3r\frac{d^3p}{(2\pi)^3}f_s(\mathbf{r},\mathbf{p},t)X(\mathbf{r},\mathbf{p})\\
&=&\frac{1}{\tilde{N}}\sum_{i=1}^{\tilde{N}}X(\mathbf{r}_i,\mathbf{p}_i).
\end{eqnarray}

We introduce a discrete time step $\Delta t$, such that during each time step the test particles propagate without colliding, following their classical trajectories. At the end of each time step collisions between them are evaluated. In a harmonic potential the trajectories are given by
\begin{eqnarray}
r_i(t_{n+1})&=&r_i(t_n)\cos(\omega_i\Delta t)+(p_i(t_n)/m\omega_i)\sin(\omega_i\Delta t), \\
p_i(t_{n+1})&=&p_i(t_n)\cos(\omega_i\Delta t)-r_i(t_n)m\omega_i\sin(\omega_i\Delta t).
\end{eqnarray}
Note that since the time step is fixed, the trigonometric functions only need to be evaluated once during the entire simulation, so that using the exact solution is more efficient than using the Verlet algorithm, as for instance in Refs.~\cite{urban, jackson}, in which case the accelerations need to be recalculated for every time step. The Verlet algorithm is more general as it is applicable for any potential. But since in this work we will only consider harmonic potentials, we will use the exact solution instead.

We evaluate the probability of a two-particle collision in the same way as described in \cite{urban}. First we must test whether a given pair of test particles reaches the point of closest approach during the present time step. This condition is important to prevent particles from attempting to collide with each other repeatedly over several consecutive time steps, an issue which will be further addressed below. If the closest approach condition is true we check if the minimal distance $d_\textnormal{min}$ of the test particles fulfils the classical condition for scattering: $\pi d_\textnormal{min}^2<\tilde{\sigma}$. If this condition is also satisfied we propose a collision at the time of closest approach. However due to Pauli statistics, even if the classical conditions for scattering are fulfilled, a collision can only take place if the new state of the particles was previously unoccupied. To take this into account we calculate the quantum mechanical scattering probability given by the Pauli term $(1-f'_s)(1-f'_{\overline{s}})$ and accept or reject the collision according to this probability. Clearly, the point-like particle picture is unsuitable for the calculation of this probability. To return to a continuous distribution we therefore have to smear out the $\delta$-functions representing the test particles, e.g.\ by Gaussians in position and momentum space:
\begin{subequations}\label{smearing}
\begin{eqnarray}
\delta(\mathbf{p}-\mathbf{p}_i)&\rightarrow&\frac{e^{-(\mathbf{p}-\mathbf{p}_i)^2/w_p^2}}{(\sqrt{\pi}w_p)^3}\\
\delta(\mathbf{r}-\mathbf{r}_i)&\rightarrow&\frac{e^{-(x-x_i)^2/w_x^2}}{\sqrt{\pi}w_x}\frac{e^{-(y-y_i)^2/w_y^2}}{\sqrt{\pi}w_y}\frac{e^{-(z-z_i)^2/w_z^2}}{\sqrt{\pi}w_z}.
\end{eqnarray}
\end{subequations}
The widths of these Gaussians, $w_x$, $w_y$, $w_z$ and $w_p$, need to be tuned so that on the one hand fluctuations due to the discrete nature of the test particle picture are smoothed out, but on the other hand the physical structure of the distribution function $f_s$ remains preserved \cite{urban}. The first condition is equivalent to
\begin{equation}
w_pw_r\gg(N/\tilde{N})^{1/3},
\label{smearingcondition1}
\end{equation}
where we introduced $w_r=(w_xw_yw_z)^{1/3}$, the geometric average of the spatial widths. The second condition implies $w_i\ll R_i$ and $w_p\ll \fermimom$. 

We also require that the smearing by the Gaussian functions preserve temperature-dependent degeneracy effects of the Fermi distribution, particularly close to the Fermi surface. This implies 
\begin{equation}
w_p\ll \fermimom(T/\fermiT)\ \ \textnormal{and}\ \ w_i\ll R_i(T/\fermiT).
\label{smearingcondition2}
\end{equation}
Note that the smearing width in momentum space is isotropic, while in position space the smearing width can be different depending on the spatial direction, if the corresponding trap frequencies are unequal. Since the Thomas-Fermi radii are inversely proportional to the corresponding trap frequencies it is sensible to choose $w_i=w_r\omega_0/\omega_i$ for the spatial widths. Furthermore Eq.~(\ref{smearingcondition2}) together with the definition of the Thomas-Fermi radii imply $w_p=m\omega_0w_r$. Hence all four smearing widths can be reduced to only one free parameter. At very low temperatures the margin given by the conditions (\ref{smearingcondition1}) and (\ref{smearingcondition2}) becomes so narrow that it is impossible to find smearing widths satisfying both conditions, without having to significantly increase the number of test particles. This limits the applicability of this setup to temperatures above approximately $0.2\fermiT$ for $\tilde{N}/N=10$. Moreover, at very low temperatures the system undergoes a phase transition into a superfluid state. This method does not include the relevant degrees of freedom of the superfluid Fermi gas and is only applicable in the normal phase.

\subsection{Auxiliary grid}
The main numerical challenge is to efficiently evaluate collisions between the test particles. The total number of pairs of opposite spin is $\tilde{N}^2/4$, which is an unfavourable scaling given that we want to use large particle numbers and a high test particle to particle ratio. In this work we develop a more efficient method than to check all possible test particle pairs. The key observation is that since the cross section is decreasing with increasing relative momentum it can never exceed $\tilde{\sigma}_\textnormal{max}=4\pi a^2N/\tilde{N}$ and consequently the maximal distance two colliding test particles can have is $d_\textnormal{max}=2a\sqrt{N/\tilde{N}}$. Having this in mind we superpose a cubic grid with cell size $d_\textnormal{max}$ on the continuous space. The grid has finite extent which can be set by demanding that a certain proportion, for instance at least $95\%$, of the test particles are within the grid. For a cigar shaped trap the grid must have larger extent in the axial direction. At the end of each time step we move systematically through all grid cells starting in one of the corners and note all possible collision partners that fulfil the classical scattering conditions. To make sure that we do not miss collisions due to boundary effects, for each particle we check not only all opposite spin particles in the same grid cell, but also in all neighbouring cells (the ones sharing a face, an edge or a vertex with the given cell). This ensures that all particles in a sphere of radius $d_\textnormal{max}$ around a given particle are definitely accounted for. This makes a total of $3^3=27$ cells for each particle, however to avoid double counting we only need to evaluate cells in the positive direction, which means on average $14$ cells per particle.

A small systematic error source remains with this setup. If the relative velocity of two particles is large they can be in non-neighbouring cells at the beginning and at the end of a time step, although in the course of the time step they come within each other's allowed collision range. Such a possible collision will then not be accounted for. However this systematic error can be minimised by choosing the time step to be sufficiently small and also by choosing the cell size to be larger than $d_\textnormal{max}$. Also note that for large relative velocities the cross section is small and collisions between very fast particles are rare events.

After having searched the entire grid for classically allowed collision pairs we proceed to choose which collisions will indeed take place. To do so we consecutively select random pairs from the list. We then propagate both particles to the point of their closest approach, let them scatter (the exact setup for determining the new positions and momenta after scattering is described in Sec.~\ref{sec:collsetup}) and then propagate them back to the original time. To account for quantum statistics we then calculate the Pauli blocking factors using the new distributions $f_s'=f_s(\mathbf{r}',\mathbf{p}')$. The continuous distributions are obtained by replacing the $\delta$-functions in (\ref{eq:fsdelta}) according to (\ref{smearing}). If we obtain a value $f_s'>1$ from the summation we set $f_s'=1$. The probability that the collision is accepted is then given by $(1-f'_s)(1-f'_{\overline{s}})$. Regardless if a collision is accepted or rejected neither of the particles concerned is allowed to collide again with another particle during the present time step. If the collision is accepted we keep the new positions and momenta. If the collision is rejected we return to the values before the collision. This procedure is repeated until all possible pairs have been evaluated.

\section{Collisional setup}
\label{sec:collsetup}
Since the picture of colliding point-like particles with well-defined positions and momenta is a classical interpretation of a quantum mechanical scattering process, it is unsurprising that there must be some ambiguity in the implementation of the collisional setup. In the semiclassical particle picture each collision has $12$ degrees of freedom: three position and three momentum components for each of the two particles, or equivalently three components of the total and relative positions and momenta. In quantum mechanics we consider wave packets rather than particles and concepts like particle position or momentum are not well-defined. Instead the system is described by operators which, depending on the symmetries of the system, commute with the Hamiltonian and define quantum numbers corresponding to conserved quantities. The concept of a trajectory does not exist, as a particle is not localised in phase space but rather smeared out over a certain phase space volume in accordance with Heisenberg's uncertainty principle. Therefore, in the collisional setup, it is not necessary to preserve for instance the exact positions of the two atoms during a collision. In the presence of an axially symmetric external potential for instance, the only conserved quantities are the total energy and the angular momentum component $L_z$.

\subsection{Angular momentum preserving setup}
\label{subsec:angmomcoll}
The method used in this work and in \cite{boltzmannletter} can be motivated as follows. As collisions are local, we can disregard for the moment the external potential and go to the centre of mass frame of the two particles. Motivated by the analogue of classical scattering we wish to conserve the total momentum, the total energy and the total angular momentum of the system during the collision. Conservation of the total momentum $\mathbf{p}=\mathbf{p}_s+\mathbf{p}_{\overline{s}}$ and the total energy $E=E_{\rm kin}=(p_s^2+p_{\overline{s}}^2)/2m$ together imply the conservation of the modulus of the relative momentum $p_{\rm rel}=|\mathbf{p}_s-\mathbf{p}_{\overline{s}}|$, since $p^2+p_{\rm rel}^2=2(p_s^2+p_{\overline{s}}^2)=4mE$. The direction of the relative momentum vector can change during the collision. Finally we must also conserve the angular momentum $\mathbf{L}=\mathbf{r}_\textnormal{rel}\times\mathbf{p}_\textnormal{rel}$. We can satisfy all these constraints by rotating both the relative momentum and relative position vectors by the same arbitrary angle around the $\mathbf{L}$-axis. The angle of this rotation is the only degree of freedom of the collision and is determined at random. From the new values $\mathbf{p}'_\textnormal{rel}$ and $\mathbf{r}'_\textnormal{rel}$ the new values of the momenta and positions of the individual particles can be recovered using total momentum conservation and the centre of mass coordinates respectively.

So far we have ignored the external potential, which would be justified if the collisions took place when the atoms were infinitesimally close to each other and therefore experiencing the same potential. However, the two colliding particles are not exactly at the same position before the collision and their relative position changes after a successful collision. Thus potential energy is not conserved and total energy conservation is not exact during a collision in general. Nevertheless, we will show below that such changes in the total energy are negligibly small for almost any collision. Furthermore, these small changes cancel each other in a many-particle system which experiences many collisions.

\subsection{Energy preserving setup}
\label{subsec:energycoll}
It is possible to preserve energy conservation exactly by employing a different setup, for instance as in \cite{urban}. In this reference, the relative position stays fixed during a collision and the relative momentum vector is rotated by a random angle in space (such a rotation has two degrees of freedom). As a direct consequence of the unrestricted rotation this setup violates angular momentum conservation. While it is true that this violation would occur naturally in non-axially symmetric potentials even for non-interacting atoms, this collisional setup provides an additional, unphysical change of angular momentum. At any rate, since in this work we consider either isotropic or axially symmetric potentials, either the total angular momentum or its axial component $L_z$ are conserved.

\section{Tests and optimal parameters}
\label{sec:boltzmanntests}
How accurately the numerical setup represents the physical picture depends crucially on the values of the simulation parameters, in particular $\tilde{N}/N$, the time step $\Delta t$ and the smearing widths $w_r$ and $w_p$. For all tests described below we use $N=10000$ atoms and $\tilde{N}/N=10$, which is sufficient for temperatures $T\geq0.2\fermiT$.

The optimal value of $\Delta t$ depends on the physical parameters of the system. Obvious requirements are that the time step must be smaller than the typical time between two collisions and that the average distance travelled during a time step must be much smaller than the diameter of the cross section, $\langle v_{\rm rel}\rangle\Delta t\ll\sqrt{\langle\tilde{\sigma}\rangle/\pi}$. Another constraint is that the time step must be smaller than the half-period with respect to the largest trap frequency, $\Delta t<\pi/\omega_{\rm max}$, but unless the aspect ratio is extremely large this condition is much weaker than the other ones. There is no lower bound on the time step, however the simulation slows down with decreasing $\Delta t$.

We first describe the tuning of the parameters and the corresponding tests of the simulation with the angular momentum preserving collisional setup \ref{subsec:angmomcoll}. We perform the same tests as in \cite{urban} to demonstrate that even with the different collisional setup we obtain very good agreement. Then we explicitly compare the different collisional setups in Sec.~{\ref{subsec:testcolls}}.

\subsection{Equilibrium collision rates}
To obtain the correct dynamical properties, for instance the damping time of excitation modes, we need to ensure that the equilibrium collision rate observed in the simulation corresponds to the correct theoretical value. The number of collisions in a given time interval can be very easily obtained from the simulation, simply by counting all test particle collisions and then multiplying by the ratio $(N/\tilde{N})$. The theoretical value for the collision rate $\gamma$ in the presence of Pauli blocking is given by the following integral,
\begin{equation}
\gamma_{\rm block}=\int d^3r\int \frac{d^3p_s}{(2\pi)^3}\frac{d^3p_{\overline{s}}}{(2\pi)^3}\int d\Omega\frac{d\sigma}{d\Omega}|\mathbf{p}_s-\mathbf{p}_{\overline{s}}|f_sf_{\overline{s}}(1-f_s')(1-f_{\overline{s}}').
\end{equation}
After inserting the Fermi-Dirac distribution this integral can be calculated numerically \cite{urban}. The numerical setup also provides a powerful tool to artificially switch off Pauli blocking. This allows us to separately check for errors related to the general numerical setup and the calculation of the Pauli blocking factors. Without Pauli blocking the integral is simpler and can be solved analytically for the Maxwell-Boltzmann distribution,
\begin{eqnarray}
\gamma_{\rm noblock}&=&\int d^3r\int \frac{d^3p_s}{(2\pi)^3}\frac{d^3p_{\overline{s}}}{(2\pi)^3}\int d\Omega\frac{d\sigma}{d\Omega}|\mathbf{p}_s-\mathbf{p}_{\overline{s}}|f_sf_{\overline{s}}\label{noblockint}\\
&=&N_\uparrow N_\downarrow\frac{2\omega_0^3}{\pi T^2}\left(1+\frac{1}{ma^2T}e^{\frac{1}{ma^2T}}\Ei\left(-\frac{1}{ma^2T}\right)\right),
\label{trapfullcollrate}
\end{eqnarray}
where $\Ei(x)=\int_{-\infty}^x(e^t/t)dt$ is the exponential integral. Furthermore we can obtain the theoretical prediction for the Pauli blocking probability by solving the integral (\ref{noblockint}) for the Fermi-Dirac distribution. The probability $p_{\rm Pauli}$ that two classically colliding fermions will indeed scatter is then given by the ratio of $\gamma_{\rm block}$ to this integral.

To find the optimal value for $\Delta t$ for each system we measure the collision rate in the absence of Pauli blocking for decreasing values of the time step and compare it to the theoretical prediction. The time step is small enough when we reach good agreement. It is important to switch off Pauli blocking for the tuning of the time step, as in the presence of Pauli blocking the collision rate is sensitive to the values of the smearing widths, which can obscure inaccuracies due to a time step that is too large. After having found the optimal time step we check that repeated (unphysical) collisions of the same particle pair are rare. This has always been found to be the case with our collisional setup. We then use this optimal value for $\Delta t$ to establish the optimal values of the smearing widths. We first identify the allowed interval for $w_p$ and $w_r$ given by the conditions (\ref{smearingcondition1}) and (\ref{smearingcondition2}) and perform the same collision rate matching as described above, this time with Pauli blocking switched on. We find that the optimal widths lie between $w_r=w_p/(m\omega_0)=1.0l_{\rm ho}$ for the lowest and $w_r=w_p/(m\omega_0)=2.0l_{\rm ho}$ for the highest temperatures used in our analysis, where $l_{\rm ho}=1/\sqrt{m\omega_0}$ is the natural harmonic oscillator length unit.

\begin{figure}[tbhp]
\includegraphics[width=.49\textwidth]{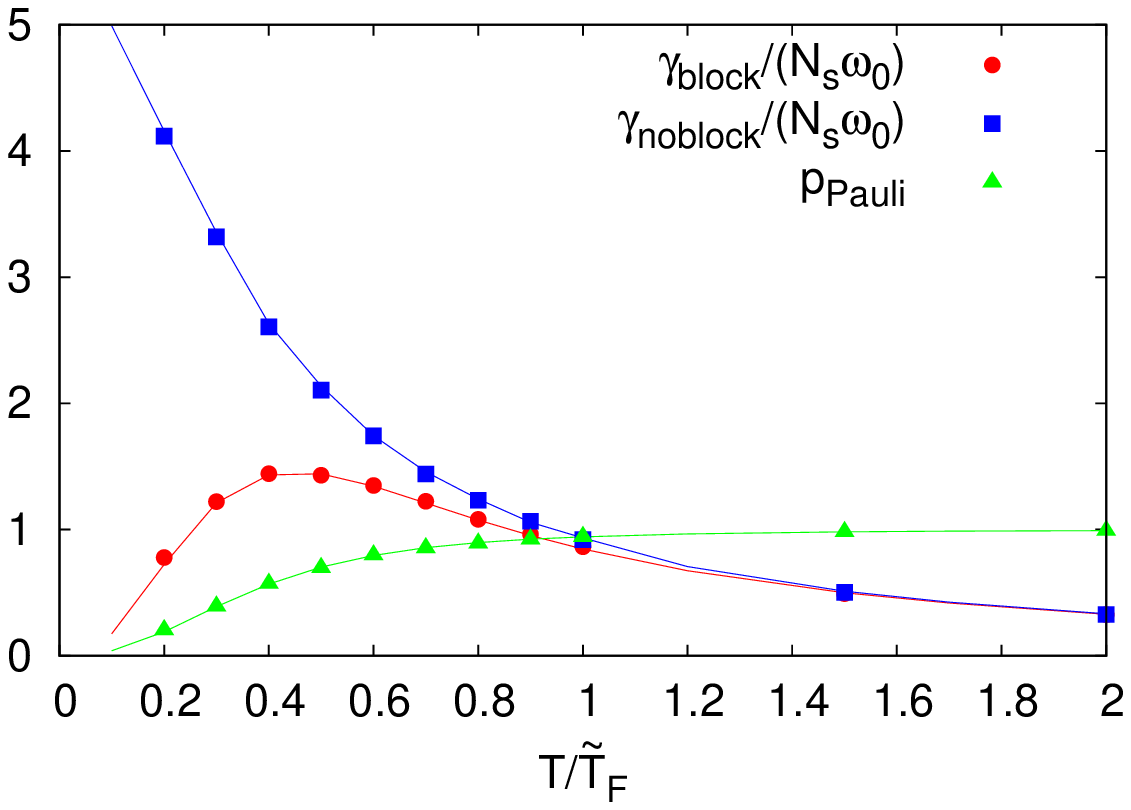}
\hfill
\includegraphics[width=.49\textwidth]{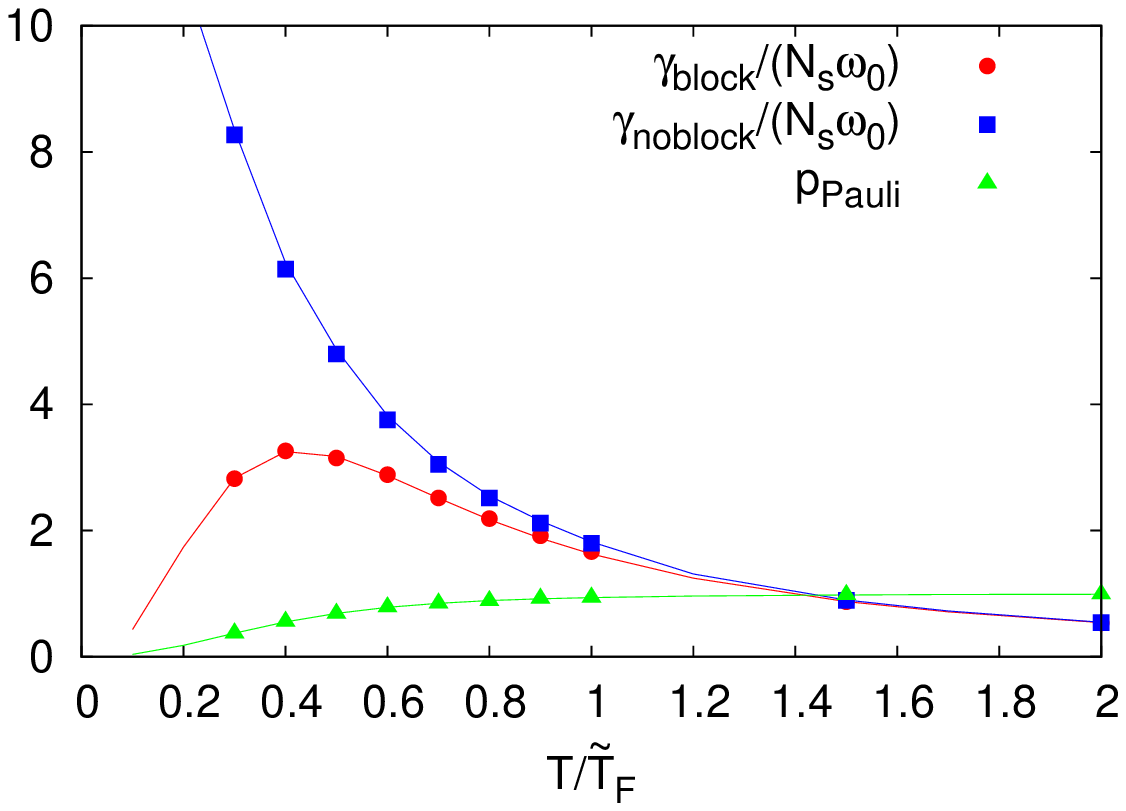}
\caption{(Colour online) The equilibrium collision rates per particle (in units of $\omega_0$) with and without Pauli blocking, as well as the Pauli probability for a successful scattering versus temperature for $|\fermimom a|=1$ (left) and $|\fermimom a|=2$ (right). The lines correspond to the theoretical prediction and the symbols to the values obtained with the simulation.\label{fig:collrates}}
\end{figure}

The measured collision rates for the optimal choice of parameters with and without Pauli blocking together with the theoretical predictions are shown in Fig.~\ref{fig:collrates}. For sufficiently high temperatures the agreement is very good. At very low temperatures it becomes increasingly difficult to resolve the conditions (\ref{smearingcondition1}) and (\ref{smearingcondition2}) for the Gaussian smearing widths simultaneously. For larger values of the scattering length this problem gets worse since test particles with a large relative distance can scatter on each other \cite{urban} and therefore the continuous distribution function needs to be resolved accurately over larger scales.

\subsection{Equilibrium energy distributions}
\begin{figure}[tbhp]
\includegraphics[width=.49\textwidth]{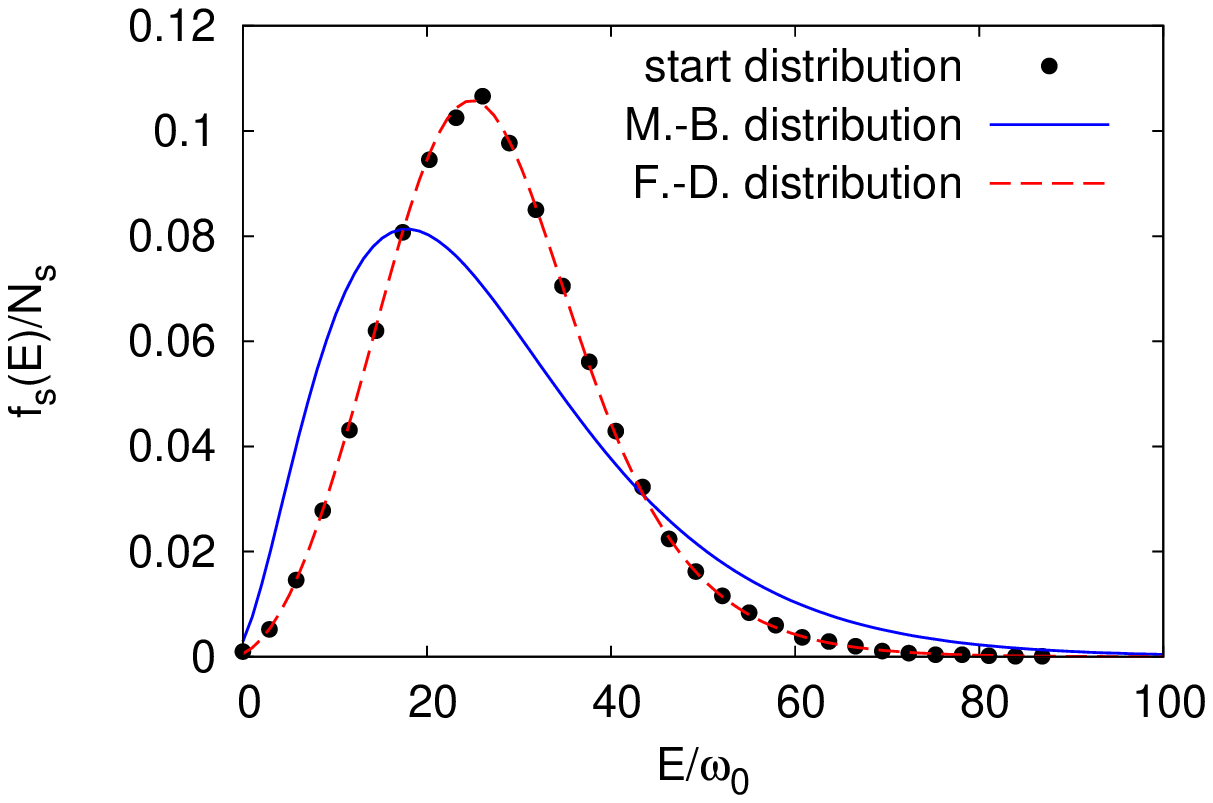}
\hfill
\includegraphics[width=.49\textwidth]{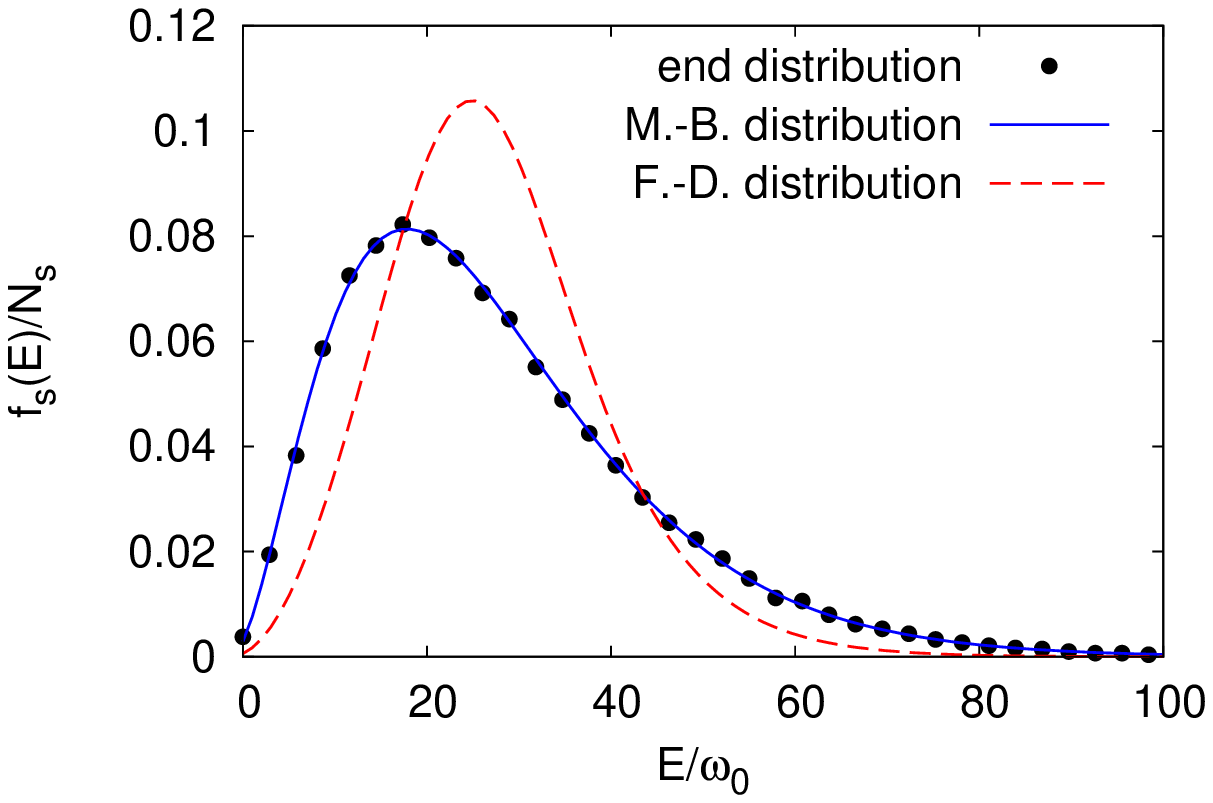}
\includegraphics[width=.49\textwidth]{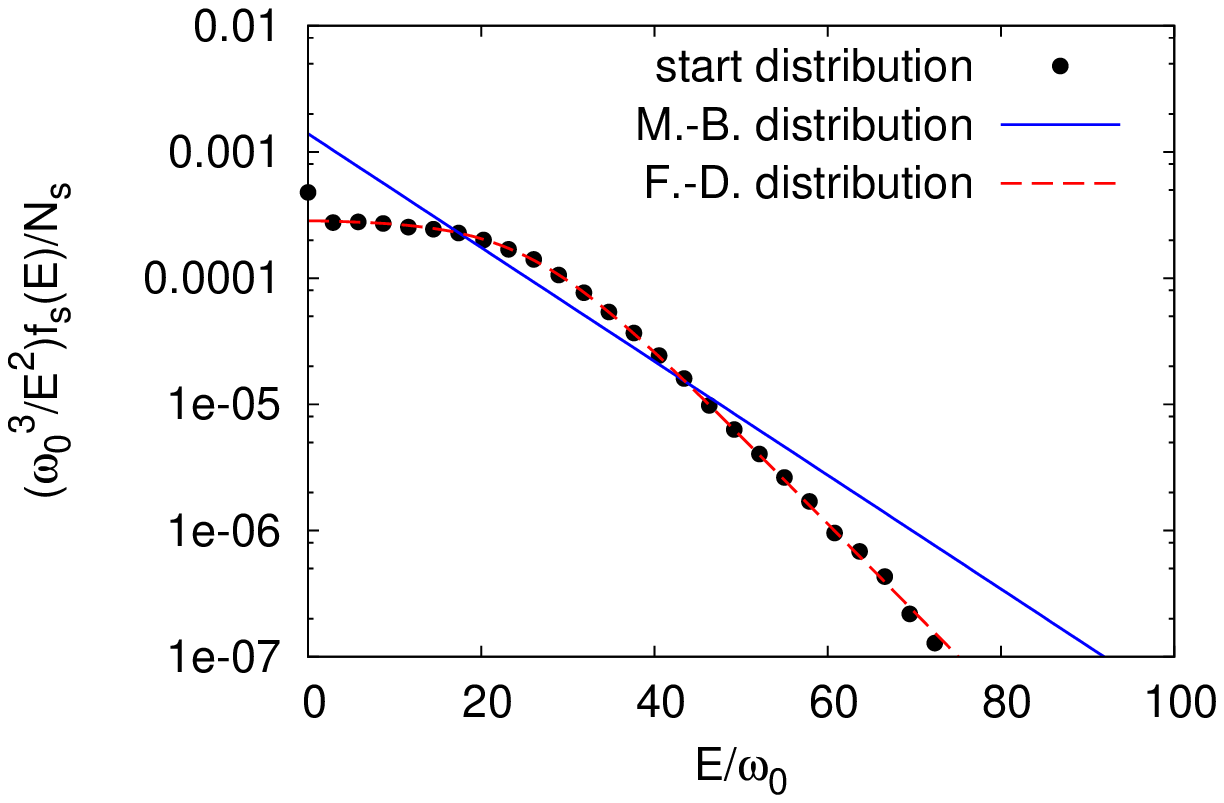}
\hfill
\includegraphics[width=.49\textwidth]{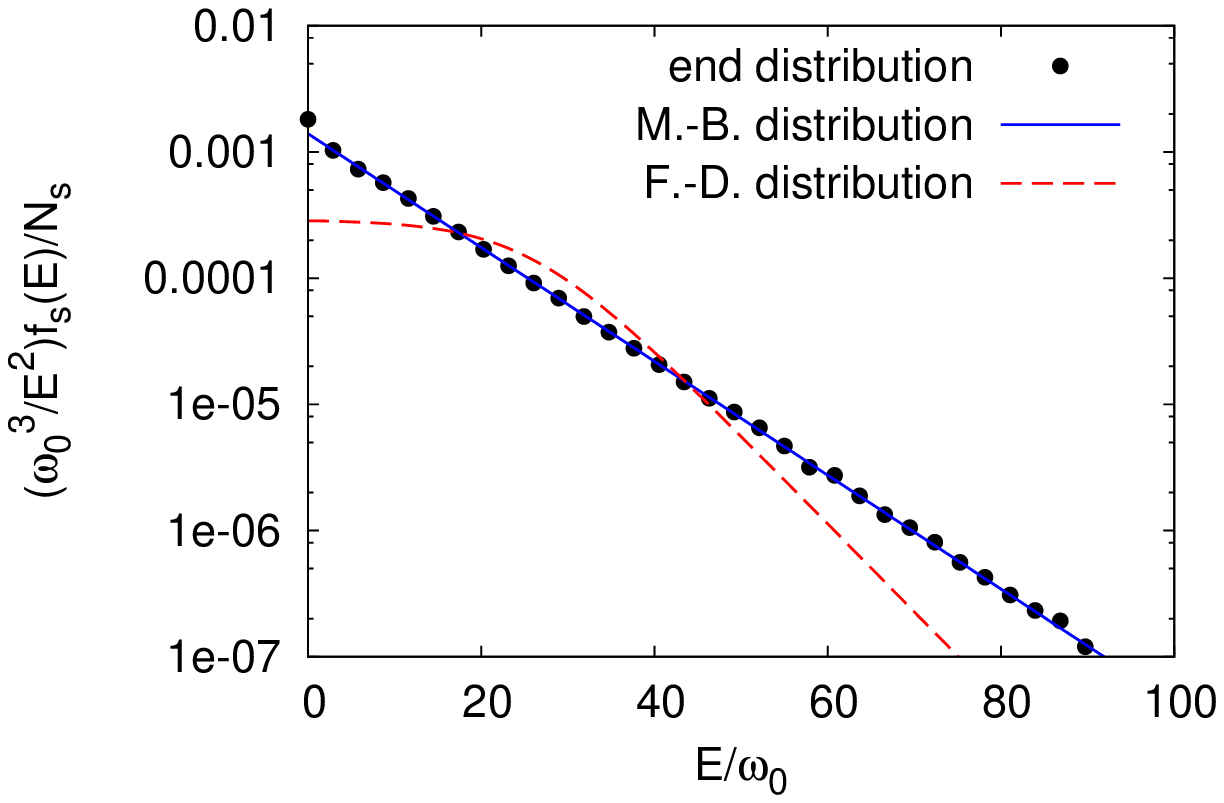}
\caption{(Colour online) The equilibrium energy distributions without Pauli blocking. The top panel shows the energy distributions on a linear scale, the bottom panel shows the energy distributions scaled by $\omega_0^3/E^2$ on a logarithmic scale. The start distribution (left) is Fermi-Dirac (F.-D.) at $T=0.2\fermiT$ and the end distribution (right) is Maxwell-Boltzmann (M.-B.) with the same average energy but at $T=0.31\fermiT$, as expected.\label{fig:FDnoblock}}
\end{figure}

\begin{figure}[tbhp]
\includegraphics[width=.49\textwidth]{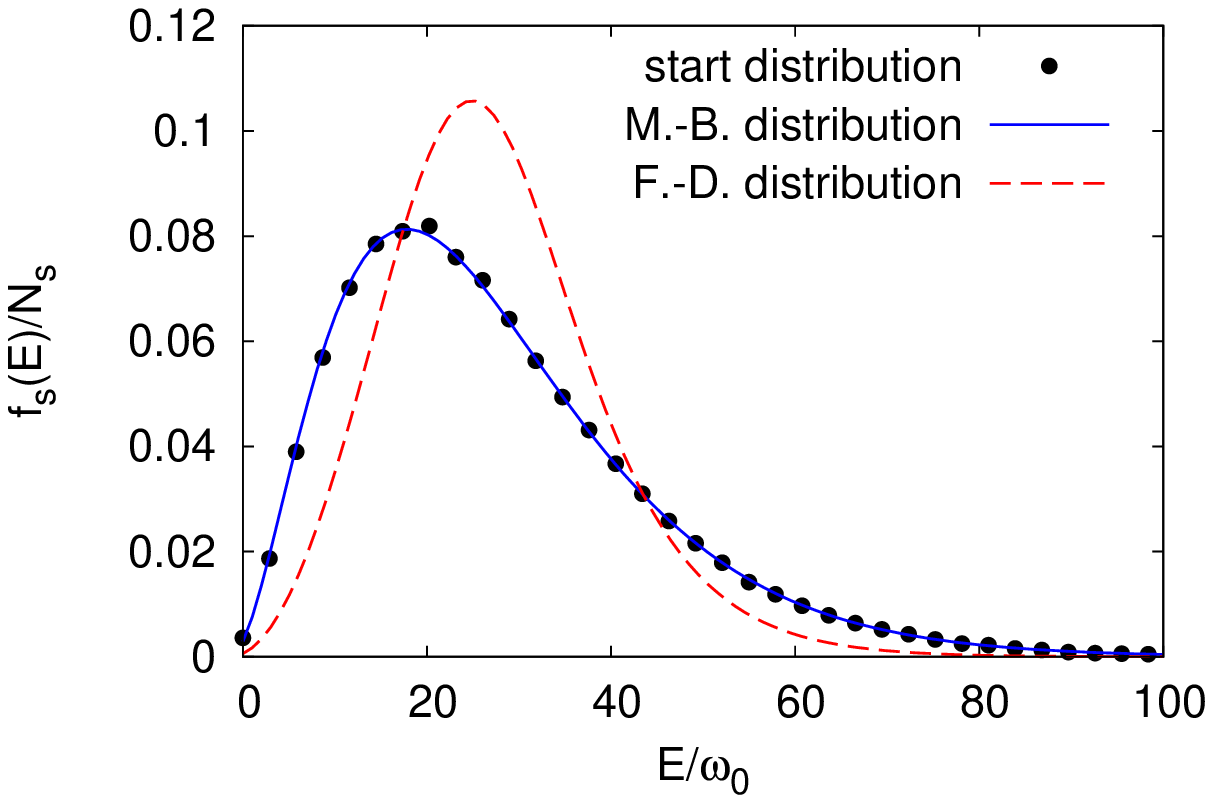}
\hfill
\includegraphics[width=.49\textwidth]{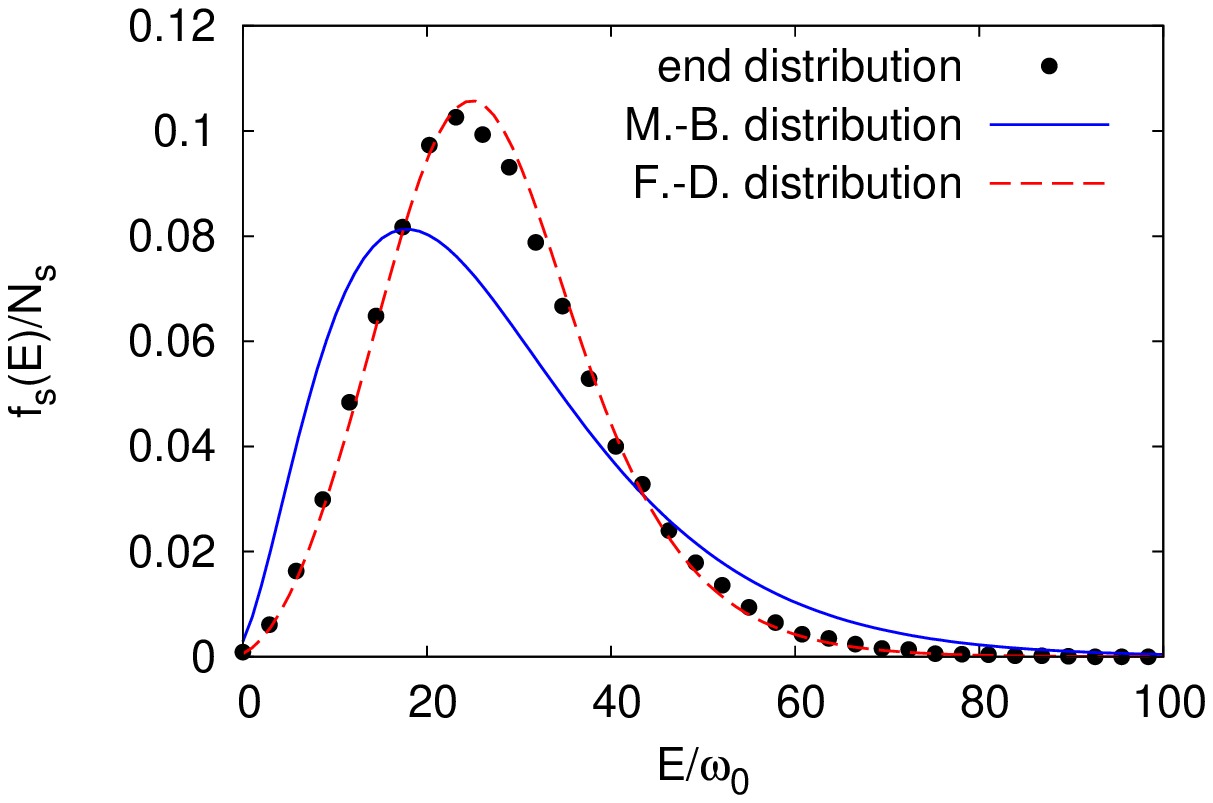}
\includegraphics[width=.49\textwidth]{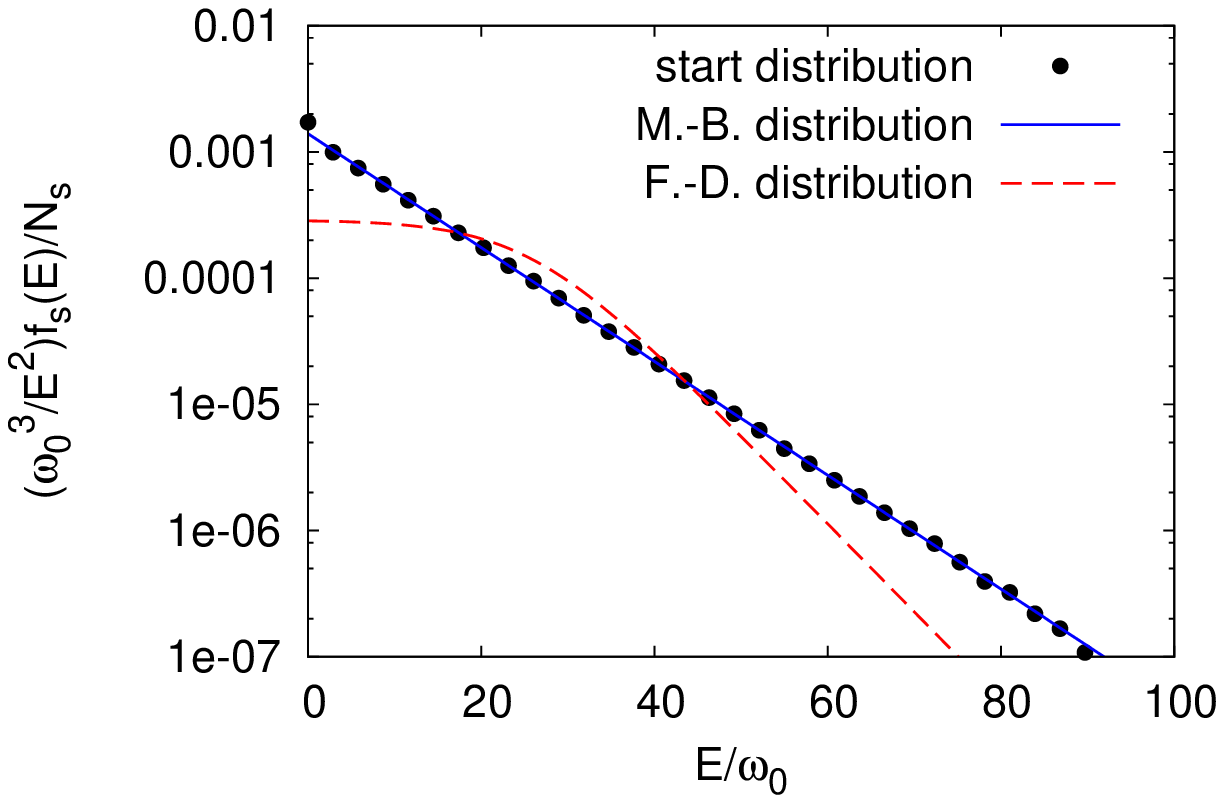}
\hfill
\includegraphics[width=.49\textwidth]{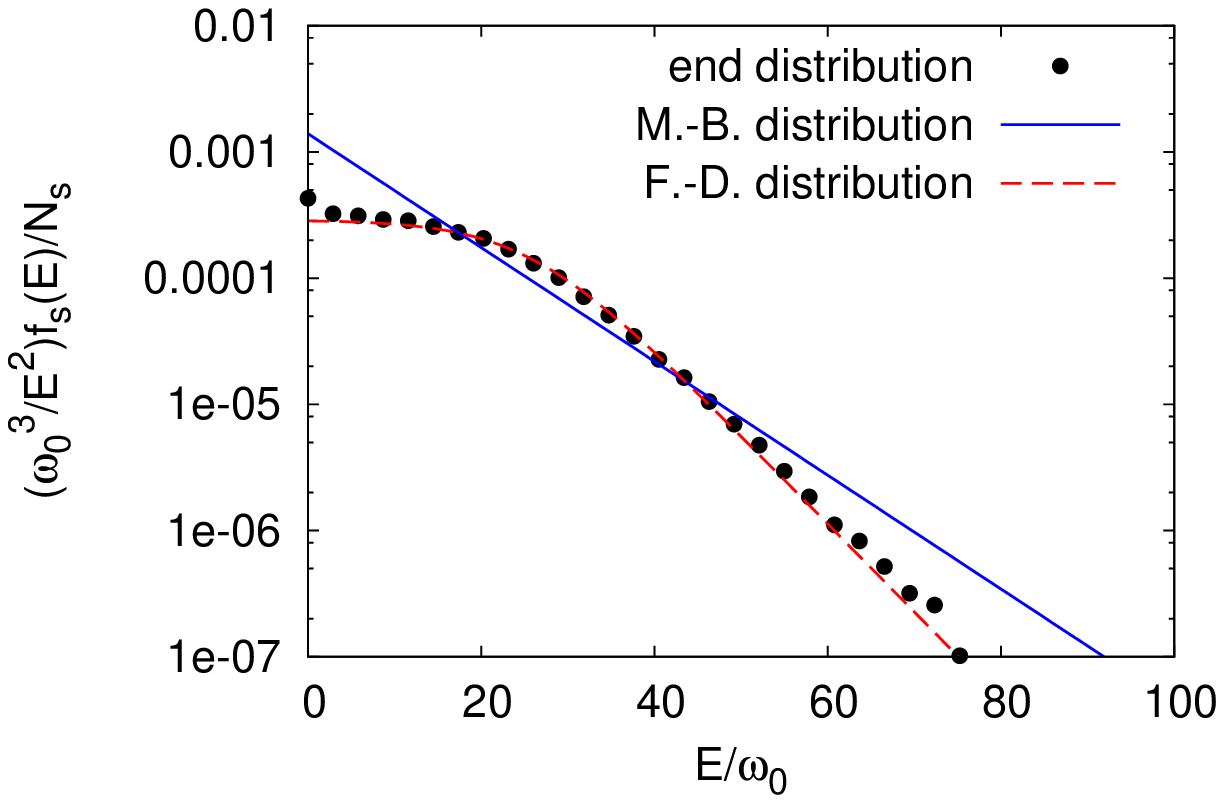}
\caption{(Colour online) The equilibrium energy distributions with Pauli blocking. The top panel shows the energy distributions on a linear scale, the bottom panel shows the energy distributions scaled by $\omega_0^3/E^2$ on a logarithmic scale. The start distribution (left) is Maxwell-Boltzmann (M.-B.) at $T=0.31\fermiT$ and the end distribution (right) is Fermi-Dirac (F.-D.) with the same average energy but at $T=0.2\fermiT$, as expected.\label{fig:MBblock}}
\end{figure}

Another important test is to check that the system thermalises to the correct equilibrium energy distribution, independently of the initial distribution. In the presence of Pauli blocking the energy is  distributed according to the Fermi-Dirac distribution, whereas without Pauli blocking the particles will be distributed according to the Maxwell-Boltzmann distribution. Figures~\ref{fig:FDnoblock} and \ref{fig:MBblock} show the results of this test for a low temperature system with $|\fermimom a|=1$ and isotropic trap frequencies. The parameter values for the time step and the smearing widths are optimal. We performed two tests of the thermalisation. First we initialised the system according to the Fermi-Dirac distribution for $T=0.2\fermiT$ (Fig.~\ref{fig:FDnoblock}) and ran the simulation without Pauli blocking. After a short time the system thermalised according to the Maxwell-Boltzmann distribution for $T=0.31\fermiT$. Note that the temperatures of the two distributions are not necessarily equal, since the equipartition theorem does not hold for the Fermi-Dirac distribution. When changing from one distribution to the other the average energy of the system remains conserved and hence the temperature of the new equilibrium state is different. Figure~\ref{fig:MBblock} shows the corresponding results for the reverse situation: the initial distribution is Maxwell-Boltzmann with $T=0.31\fermiT$ and Pauli blocking is switched on. It is clearly visible from both figures that the correct equilibrium distribution is always attained at the end of the simulation. This agreement improves further at higher temperatures.

\subsection{Collective excitations}
Collective excitations emerge when a many-particle system is perturbed away from equilibrium. Here we will discuss three different excitation modes, the sloshing mode (also known as dipole or Kohn mode), the breathing mode (monopole mode) and the quadrupole mode. We will confirm that the simulation gives the correct frequencies and damping properties of these modes. The following tests were performed for a spherical trap $\omega_x=\omega_y=\omega_z=\omega_0$.

\begin{figure}
\centering
\includegraphics[width=0.6\textwidth]{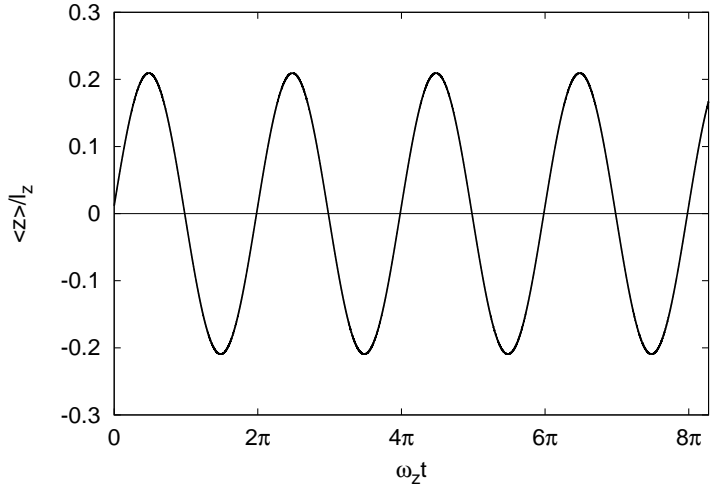}
\caption{Simulation of the equilibrium sloshing mode $\langle z\rangle/l_z$, where $l_z=1/\sqrt{m\omega_z}$. The mode is undamped and the oscillation frequency equals $\omega_z$.\label{fig:sloshingmode}}
\end{figure}
The sloshing mode is excited by a small displacement of the centre of mass from its equilibrium position, or equivalently by a short-lived force represented by an additional linear term in the potential. The time evolution of each of the three centre of mass coordinates $\langle r_i\rangle$ is well-known to be an undamped oscillation with the corresponding harmonic oscillator frequency $\omega_i$. Figure~\ref{fig:sloshingmode} shows such an oscillation for a system at $|\fermimom a|=1$ and $T=0.2\fermiT$.

\begin{figure}
\centering
\includegraphics[width=0.6\textwidth]{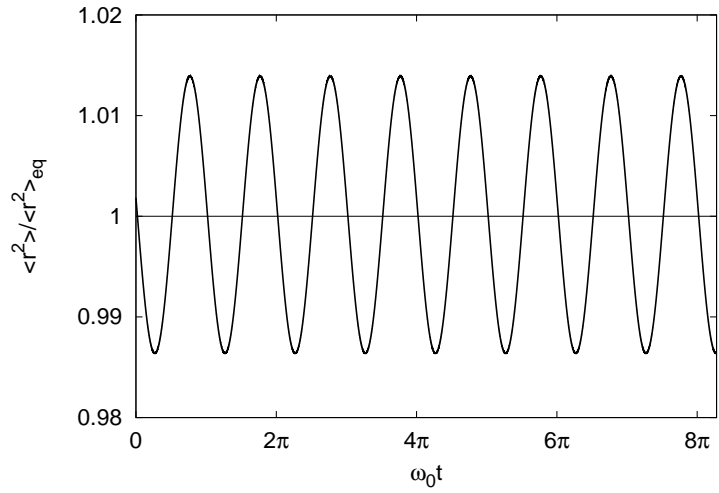}
\caption{Simulation of the normalised equilibrium breathing mode $\langle r^2\rangle/\langle r^2\rangle_{\rm eq}$. The breathing mode is undamped and the oscillation frequency equals $2\omega_z$.\label{fig:breathingmode}}
\end{figure}
The breathing mode can be excited by compressing or expanding the cloud. In a spherical trap this yields an undamped oscillation of $\langle r^2\rangle$ with frequency $2\omega_0$ around its equilibrium value \cite{guery1999collective}. Figure~\ref{fig:breathingmode} illustrates this mode for a system with $|\fermimom a|=1$ and $T=0.2\fermiT$.

\begin{figure}[tbhp]
\includegraphics[width=.49\textwidth]{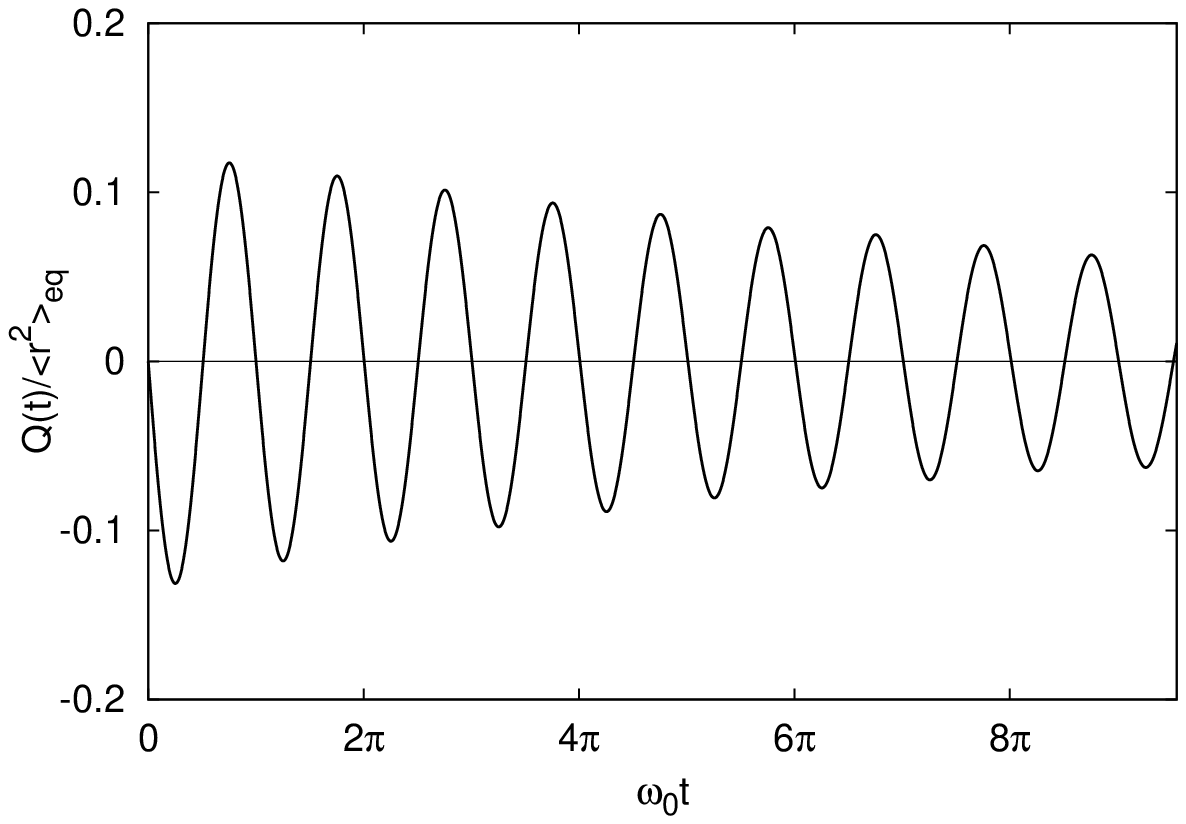}
\hfill
\includegraphics[width=.49\textwidth]{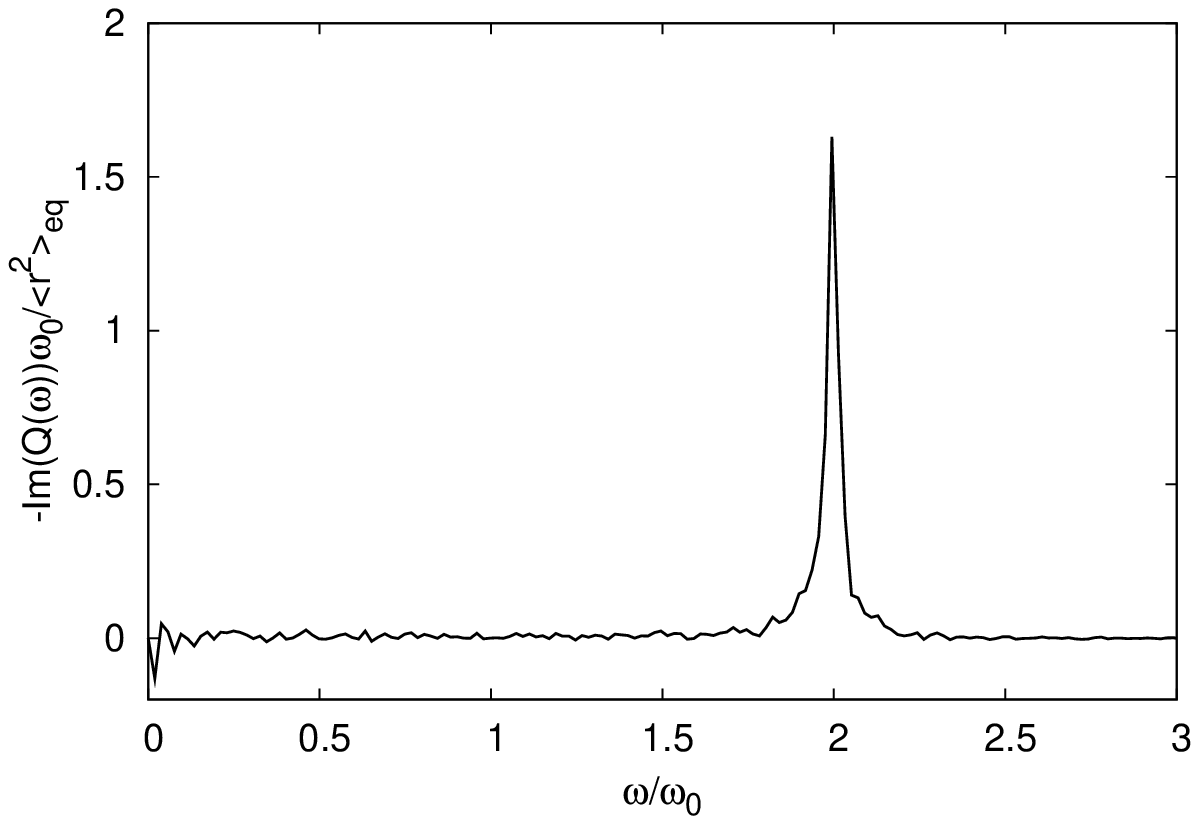}
\includegraphics[width=.49\textwidth]{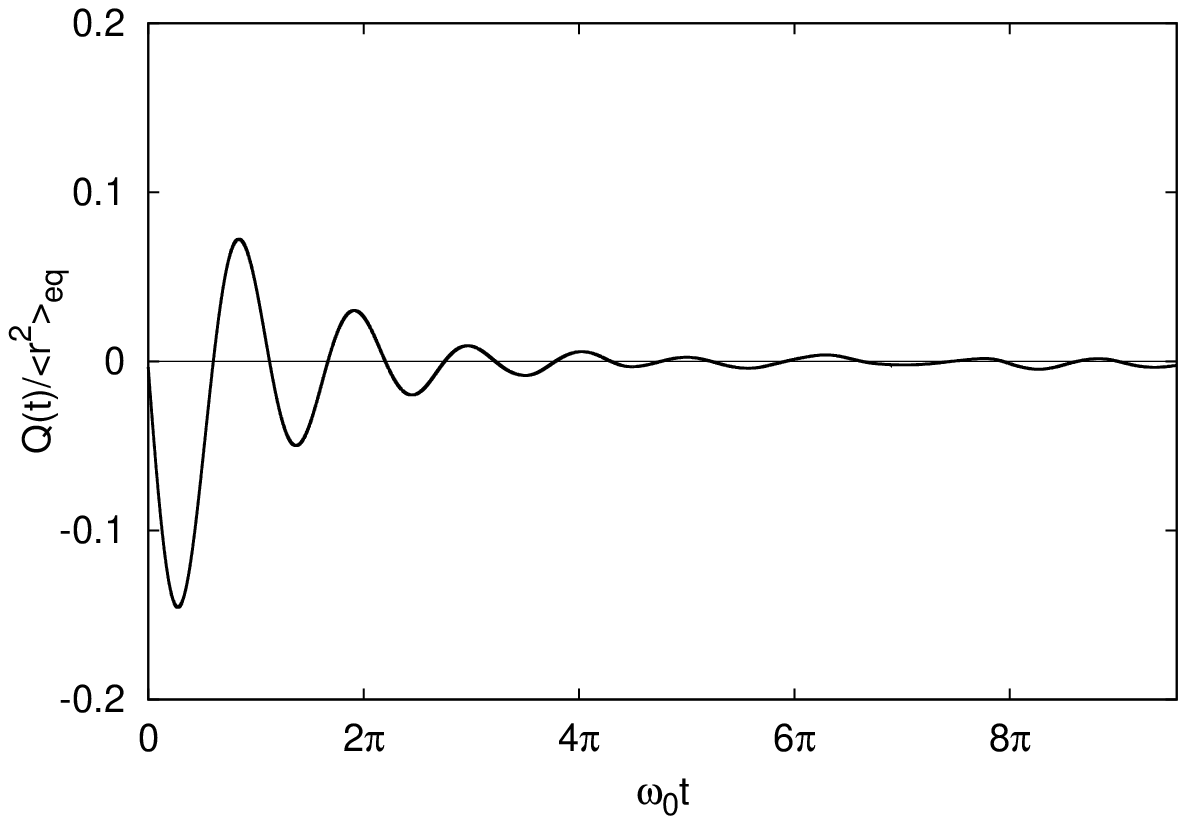}
\hfill
\includegraphics[width=.49\textwidth]{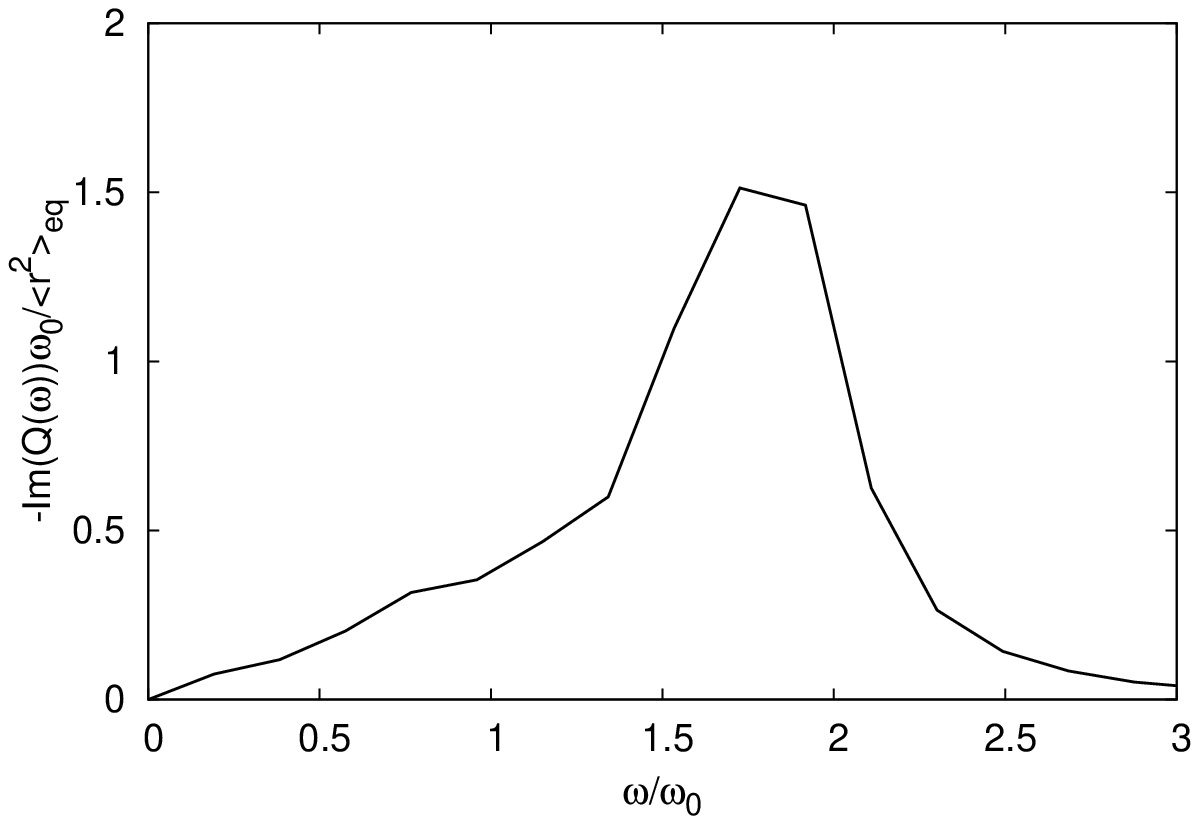}
\caption{Left: Simulation of the normalised equilibrium quadrupole mode $Q(t)/\langle r^2\rangle_{\rm eq}$ at high temperature $T=2\fermiT$ (top panel) and at low temperature $T=0.4\fermiT$ (bottom panel). Right: The imaginary part of the corresponding Fourier transforms of the numerical data gives information about the oscillation frequency and the damping.\label{fig:quadrupolemode}}
\end{figure}
By perturbing the system via a short-lived small increase in one or several of the trap frequencies we can excite the quadrupole mode. We excite the quadrupole mode $Q(t)=\langle x^2\rangle-\langle y^2\rangle$ by applying the perturbation $\Delta p_x=-cx$ and $\Delta p_y=cy$ with $c=0.2m\omega_0$ in the same way as in \cite{urban}. Unlike the sloshing and the breathing mode this mode is damped. The frequency of the quadrupole mode at high temperatures approaches the ideal gas value $2\omega_0$, while at low temperatures it is closer to the hydrodynamic frequency $\sqrt{2}\omega_0$. Figure~\ref{fig:quadrupolemode} shows the quadrupole mode for $|\fermimom a|=1$ in the high and in the low temperature regime. In both cases the damping of the mode is clearly visible. The frequency of the oscillation can be extracted from the corresponding Fourier transform $Q(\omega)=\int_0^\infty dt Q(t)e^{i\omega t}$ and is in agreement with the theoretical prediction while the damping can be extracted from its imaginary part. We see that the damping is stronger at the lower temperature $T=0.4\fermiT$ when collisions are more frequent (see Fig.~\ref{fig:collrates} for a plot of the collision rate). 

\subsection{Comparison of different collisional setups}
\label{subsec:testcolls}
\begin{figure}[thbp]
\centering
\includegraphics[width=0.6\textwidth]{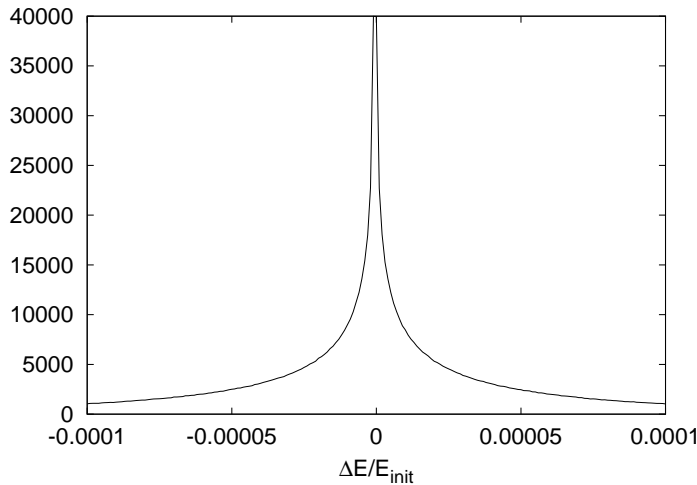}
\caption{\label{fig:energyhist}Probability density (obtained from the corresponding normalised histogram) for the relative change of the energy of a colliding particle pair $\Delta E/E_\textnormal{init}=(E_\textnormal{final}-E_\textnormal{init})/E_\textnormal{init}$ for $|\fermimom a|=1$, $T=0.4\fermiT$, $\omega_\perp/\omega_z=100$ and $\tilde{N}=10N=100000$.}
\end{figure}

To study the impact of the collisional setup on the outcome of the simulation we have implemented the setups \ref{subsec:angmomcoll} (angular momentum preserving) and \ref{subsec:energycoll} (energy preserving) and confirmed that they generate the same results for both equilibrium and non-equilibrium systems. First we analyse the magnitude of the changes in the total energy with the angular momentum conserving setup. Figure~\ref{fig:energyhist} shows a typical histogram of the relative change of the energy of a particle pair $\Delta E/E_\textnormal{init}=(E_\textnormal{final}-E_\textnormal{init})/E_\textnormal{init}$ after a successful collision. The histogram is sharply peaked around zero. In this typical example the majority of collisions (approximately 84\%) conserve the energy of the colliding pair with an accuracy of up to $|\Delta E/E_\textnormal{init}|\leq10^{-4}$. Since collisions are frequent and the energy changes have random sign we also observe large cancellation effects, such that the total energy of the system is conserved with an accuracy of the order of $10^{-7}$ at any given time. This is smaller than for instance the energy deviations due to Verlet's algorithm quoted in \cite{urban}.

\begin{figure}[thbp]
\centering
\includegraphics[width=0.6\textwidth]{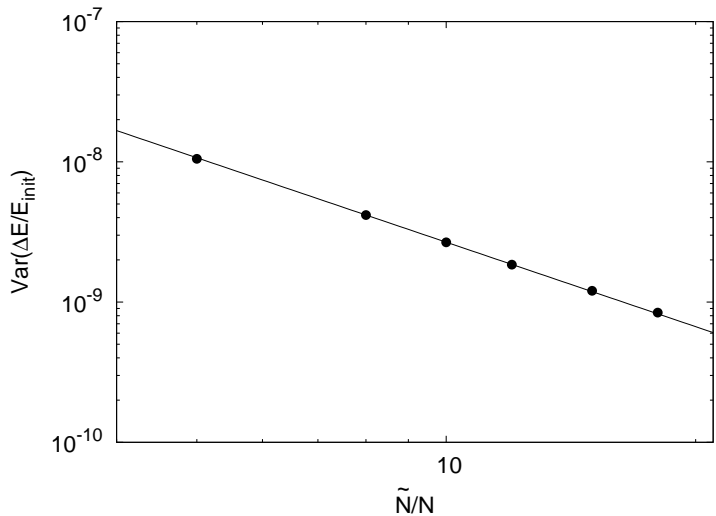}
\caption{\label{fig:variancetestpart}Double-logarithmic plot of the variance of the energy change during a collision $\textnormal{Var}(\Delta E/E_\textnormal{init})$ for $|\fermimom a|=1$, $T=\fermiT$ and $\omega_\perp/\omega_z=100$ as a function of the test particle to particle ratio $\tilde{N}/N$. The line corresponds to a linear fit with constant slope $-2$ on the double-logarithmic scale.}
\end{figure}

To understand the good accuracy of energy conservation we also make a qualitative theoretical estimate. During the collision the kinetic energy is conserved, such that it is sufficient to consider the potential energy $V$ of the two particles,
\begin{equation}
V=\frac{m}{2}\left[\omega_\perp^2(x_1^2+y_1^2+x_2^2+y_2^2)+\omega_z^2(z_1^2+z_2^2)\right],
\end{equation}
where $(x_i,y_i,z_i)$ are the coordinates of the particle $i$ and $\omega_\perp=\omega_x=\omega_y$ is the radial trap frequency. We can recast this expression in terms of the centre of mass and the relative coordinates $(X,Y,Z)$ and $(x,y,z)$ respectively,
\begin{equation}
V=m\left[\omega_\perp^2(X^2+Y^2)+\omega_z^2 Z^2\right]+\frac{m}{4}\left[\omega_\perp^2(x^2+y^2)+\omega_z^2z^2\right].
\end{equation}
The position of the centre of mass is not affected by the collision. The variation of energy is thus only due to the contribution of the relative motion $V_{\rm rel}$ defined by 
\begin{equation}
V_{\rm rel}=\frac{m}{4}\left[\omega_\perp^2(x^2+y^2)+\omega_z^2z^2\right]
\end{equation}
We now define a second frame $(x',y',z')$ such that the $z'$-axis is parallel to the relative angular momentum $\bm L$ of the pair. In addition, since the trap is rotationally invariant around $z$, we can choose the initial coordinate system such that $\bm L$ is in the $(x,z)$ plane, in which case $y'=y$. If $\theta$ is the angle between $\bm L$ and the $z$ axis, the two sets of coordinates are related by the following relations 
\begin{eqnarray*}
x&=&x'\cos\theta+z'\sin\theta\\
y&=&y'\\
z&=&z'\cos\theta-x'\sin\theta.
\end{eqnarray*}
By conservation of angular momentum, the relative coordinates evolve in the $(x',y')$ plane. We thus have $z'=0$, $x'=d\cos\varphi$ and $y'=d\sin\varphi$, where $d$ is the relative distance between the two particles. In terms of $d$ and $\varphi$, the relative potential energy is
\begin{equation}
V_{\rm rel}=\frac{m\omega_\perp^2 d^2}{4}+\frac{m d^2}{4}(\omega_z^2-\omega_\perp^2)\sin^2\theta\cos^2\varphi. 
\end{equation}
During the collision $d$ and $\theta$ are conserved. The variation of energy is thus only due to the rotation in the $(x',y')$ and we obtain
\begin{equation}
\Delta E=\frac{m d^2}{4}(\omega_z^2-\omega_\perp^2)\sin^2\theta\left[\cos^2\varphi_{\rm f}-\cos^2\varphi_{\rm i}\right],
\end{equation} 
where $\varphi_{\rm i}$ and $\varphi_{\rm f}$ correspond to the initial and final values of $\varphi$ respectively. As an approximation, we normalise to the average total energy of the pair, $\bar E\approx6T$, and obtain
\begin{equation}
\frac{\Delta E}{\bar E}=\frac{m d^2}{24T}(\omega_z^2-\omega_\perp^2)\sin^2\theta\left[\cos^2\varphi_{\rm f}-\cos^2\varphi_{\rm i}\right].\label{eq:envariation}
\end{equation} 
As shown above the distance between the particles can be at most $d_\textnormal{max}=2a\sqrt{N/\tilde{N}}$. Hence for the range of parameters considered in this study the constant prefactor is of order $\lesssim10^{-4}$. The trigonometric functions further reduce the variation of the energy and  are responsible for the pointed shape of the histogram. Note that the method of test particles is important for a small energy variation. Due to $\tilde{N}$ being larger than $N$ the cross section and hence the average distance between two scattering partners is reduced, which leads to a smaller $\Delta E/E_\textnormal{init}$ (in our case by a factor of $\tilde{N}/N=10$). To demonstrate this effect Fig.~\ref{fig:variancetestpart} shows the variance of the energy change as a function of the ratio $\tilde{N}/N$. The data is well-fitted by a line with slope $-2$ on a double-logarithmic scale, which confirms the relation (\ref{eq:envariation}). Analogous scaling laws can be shown also for the dependence on the temperature and the scattering length.

We also compared the time evolutions of the centre of mass coordinates in response to a perturbation. The differences between the two setups were found to be smaller than the statistical fluctuations. Taking these results into account we conclude that the two setups \ref{subsec:angmomcoll} and \ref{subsec:energycoll} are equivalent on a macroscopical scale and we have the freedom to choose whichever option appears more convenient.

We adopt the angular momentum conserving setup \ref{subsec:angmomcoll}, not only because of this property, but also since the setup from \cite{urban} has a small technical disadvantage. Since the new direction of the relative momentum vector is chosen uniformly on a sphere, in many cases the particles are found to approach each other again after a successful collision. This implies that in the following time step they are likely to undergo another collision. The proportion of such events compared to the total number of collisions can reach up to $40\%$ at the lowest temperature considered in this study and amounts to around $10\%$ at $T=\fermiT$. To avoid this overcounting of collisions one needs to implement an additional routine that prevents particles from colliding with each other repeatedly within short time intervals. In other words the molecular chaos assumption does not hold with the setup from \cite{urban} and needs to be enforced artificially. In our setup repeated scattering is so rare that this effect can be neglected. This is because the relative position and the relative momentum vectors of the particle pair are rotated by the same random angle and hence their relative angle is preserved. After a collision takes place the particles continue to move on trajectories that diverge from each other.

\subsection{Outlook}
In addition to the two collisional setups described above we propose a third very general collisional setup which can be explored in future work.

To preserve macroscopical averages we keep the centre of mass coordinates and momentum fixed and work with the six degrees of freedom for the relative momentum and position. Based on the idea of a delocalised particle pair, we assume that the relative phase space coordinates of the particles are distributed according to a Gaussian. We then choose the new relative position and momentum randomly according to the Gaussian probability distribution with the additional constraints given by the symmetries of the problem. Energy and $L_z$ conservation for instance reduce the problem by two degrees of freedom. If energy and total angular momentum are conserved the problem is reduced by four degrees of freedom and so on.

It is important to stress however, that the task of defining a probability measure on a complicated hypersurface in the phase space (in our case defined by the constraints of energy and angular momentum conservation) is a mathematically highly nontrivial problem and will pose a challenge for the numerical implementation.

\section{Conclusion}
\label{sec:summary}
We have implemented a Boltzmann equation approach to the simulation of the semiclassical regime of two-component Fermi gases. We described how to implement the method using test particles and stressed the advantages of the auxiliary grid. We also presented an extensive range of tests for the simulation. The main focus of this work was a discussion of the different possibilities for implementing collisions, namely the angular momentum conserving and the energy conserving setups. We studied in detail the error in the energy of the former scheme and showed that it is indeed extremely small, so that it becomes a valid alternative way of calculating the collisional integral and has distinct advantages since it minimises repeated scattering events. In future work we intend to apply the method to systems with components of unequal mass which bring a greater variety of dynamical regimes and which can provide a more stringent test of the Boltzmann equation approach to these systems.

\section*{Acknowledgements}
We thank A.~Sommer, M.~Urban and M.~Zwierlein for useful discussions. This work has made use of the resources provided by the Cambridge HPC Facility. O.G.\ acknowledges support from the German Academic Exchange Service (DAAD) and the Excellence Cluster ``Nanosystems Initiative Munich (NIM)''. F.C.\ acknowledges support from ERC (project FERLODIM), R\'egion Ile de France (IFRAF) and Institut Universitaire de France. C.L.\ acknowledges support from the EPSRC through the Advanced Fellowship EP/E053033/1.

\section*{References}
\bibliographystyle{elsarticle-num}
\bibliography{longboltzmann}

\begin{thebibliography}{10}
\expandafter\ifx\csname url\endcsname\relax
  \def\url#1{\texttt{#1}}\fi
\expandafter\ifx\csname urlprefix\endcsname\relax\def\urlprefix{URL }\fi
\expandafter\ifx\csname href\endcsname\relax
  \def\href#1#2{#2} \def\path#1{#1}\fi

\bibitem{inguscio2006ultracold}
M.~Inguscio, W.~Ketterle, C.~Salomon (Eds.), {Proceedings of the International
  School of Physics ``Enrico Fermi'' on Ultracold Fermi gases, Course CLXIV,
  Varenna}, Societ\`a Italiana di Fisica, 2006.

\bibitem{Zwierlein}
A.~{Sommer}, M.~{Ku}, G.~{Roati}, M.~W. {Zwierlein}, {Universal spin transport
  in a strongly interacting Fermi gas}, Nature 472 (2011) 201--204.
\newblock \href {http://dx.doi.org/10.1038/nature09989}
  {\path{doi:10.1038/nature09989}}.

\bibitem{ZwierleinImbalanced}
A.~{Sommer}, M.~{Ku}, M.~W. {Zwierlein}, {Spin transport in polaronic and
  superfluid Fermi gases}, New~J.~Phys. 13~(5) (2011) 055009.
\newblock \href {http://dx.doi.org/10.1088/1367-2630/13/5/055009}
  {\path{doi:10.1088/1367-2630/13/5/055009}}.

\bibitem{Trenkwalder}
A.~Trenkwalder, C.~Kohstall, M.~Zaccanti, D.~Naik, A.~I. Sidorov, F.~Schreck,
  R.~Grimm, Hydrodynamic expansion of a strongly interacting fermi-fermi
  mixture, Phys.~Rev.~Lett. 106~(11) (2011) 115304.
\newblock \href {http://dx.doi.org/10.1103/PhysRevLett.106.115304}
  {\path{doi:10.1103/PhysRevLett.106.115304}}.

\bibitem{Thomas}
C.~Cao, E.~Elliott, J.~Joseph, H.~Wu, J.~Petricka, T.~SchŠfer, J.~E. Thomas,
  Universal quantum viscosity in a unitary fermi gas, Science 331 (2010) 58--.
\newblock \href {http://dx.doi.org/10.1126/science.1195219}
  {\path{doi:10.1126/science.1195219}}.

\bibitem{boltzmannletter}
O.~Goulko, F.~Chevy, C.~Lobo, Collision of two spin-polarized fermionic clouds,
  Phys.~Rev.~A 84 (2011) 051605.
\newblock \href {http://dx.doi.org/10.1103/PhysRevA.84.051605}
  {\path{doi:10.1103/PhysRevA.84.051605}}.

\bibitem{urban}
T.~Lepers, D.~Davesne, S.~Chiacchiera, M.~Urban, Numerical solution of the
  boltzmann equation for the collective modes of trapped fermi gases,
  Phys.~Rev.~A 82~(2) (2010) 023609.
\newblock \href {http://dx.doi.org/10.1103/PhysRevA.82.023609}
  {\path{doi:10.1103/PhysRevA.82.023609}}.

\bibitem{mythesis}
O.~Goulko, {Thermodynamic and hydrodynamic behaviour of interacting Fermi
  gases}, Ph.D. thesis, University of Cambridge (2011).

\bibitem{jackson}
B.~Jackson, E.~Zaremba, Modeling bose-einstein condensed gases at finite
  temperatures with n-body simulations, Phys.~Rev.~A 66~(3) (2002) 033606.
\newblock \href {http://dx.doi.org/10.1103/PhysRevA.66.033606}
  {\path{doi:10.1103/PhysRevA.66.033606}}.

\bibitem{wade}
A.~C.~J. Wade, D.~Baillie, P.~B. Blakie, {Direct simulation Monte Carlo method
  for cold-atom dynamics: Classical Boltzmann equation in the quantum collision
  regime}, Phys.~Rev.~A 84~(2) (2011) 023612.
\newblock \href {http://dx.doi.org/10.1103/PhysRevA.84.023612}
  {\path{doi:10.1103/PhysRevA.84.023612}}.

\bibitem{guery1999collective}
D.~Gu\'ery-Odelin, F.~Zambelli, J.~Dalibard, S.~Stringari, Collective
  oscillations of a classical gas confined in harmonic traps, Phys.~Rev.~A
  60~(6) (1999) 4851--4856.
\newblock \href {http://dx.doi.org/10.1103/PhysRevA.60.4851}
  {\path{doi:10.1103/PhysRevA.60.4851}}.

\end{thebibliography}
\end{document}